# Tuning and Suppression of YIG Magnetisation Dynamics via Antiferromagnetic Interface Coupling

Oscar Cespedes,[1,*] Hari B. Vasili,[1] Matthew Rogers,[1] Paul S. Keatley,[2] Manan Ali,[1] Bryan Hickey,[1] and Robert J. Hicken[2]

[1]*School of Physics and Astronomy, University of Leeds, Leeds, UK*

[2]*Department of Physics and Astronomy, University of Exeter, Exeter, UK*

*\*o.cespedes@leeds.ac.uk*

**The magnetisation dynamics of yttrium iron garnet ($Y_3Fe_5O_{12}$, YIG) are key to the operation of spintronic and microwave devices. Here, we report a pathway to manipulate the frequency, damping and absorption of YIG thin films via interface coupling. The growth on YIG of PtMn, a metallic antiferromagnet, leads to a power dependence of the oscillation frequency and an increased linewidth at low fields. In gadolinium iron garnet/YIG film bilayers, the two films couple antiferromagnetically at low temperatures and there is a strong damping of the magnetisation dynamics that is further enhanced at the spin-flop field, suppressing the FMR signal. When combining both GdIG and PtMn interfaces, we can tune the exponent of the power dependence of frequency with field and achieve an almost complete quenching of the magnetisation dynamics over a range of fields/frequencies due to non-collinear magnetic order. These effects offer a means to tune and suppress magnetisation dynamics for frequency filters, magnonics, spin pumping and other applications.**

Ferromagnetic resonance (FMR), where magnetisation dynamics are excited by a microwave field, depends on the magnetic field and crystal orientation.[1] When a field $H_{app}$ is applied out-of-plane (OOP) to a thin film, the FMR has a characteristic frequency $f$ proportional $H_{app}$, the gyromagnetic ratio $\gamma$ and the effective internal field $H_{eff}$, which includes the anisotropy and demagnetisation;[2]

$$f = \frac{\gamma}{2\pi}(H_{app} - H_{eff}) \qquad (1)$$

High quality epitaxial, insulating and ferrimagnetic YIG thin films are commonly grown on gadolinium gallium garnet substrates ($Gd_3Ga_5O_{12}$; GGG) due to a lattice mismatch of only 0.06% in the (111) orientation. The FMR damping coefficient, $\alpha$, is usually higher in thin films than in bulk, but coefficients of $10^{-5}$ - $10^{-4}$ can be achieved in films grown by pulsed laser deposition and sputtering depending on thickness and quality.[3-6] This low damping makes YIG a prototypical material for applications in microwave filters, isolators, circulators and resonators, and also in magnonics and spintronics.[7-12] Here, we RF sputter 60-70 nm thick YIG films onto GGG substrates with optimised oxygen pressure, showing a room temperature damping of the order of $2 \times 10^{-4}$ and a coercivity of less than 0.1 mT. The GGG substrate is paramagnetic with a linear dependence of magnetisation on field. At low temperatures when the field is applied OOP, it is very difficult to disentangle the linear magnetic responses of both the GGG substrate and the YIG film because the magnetic susceptibility of the YIG thin film is orders of magnitude lower than that of the bulk GGG substrate. However, the GGG substrate does not contribute to the FMR signal and therefore OOP dynamic characterisation of YIG films is possible at 10 K. Adding a 5 nm thick PtMn layer on top of YIG has a small effect on the damping, probably due to the transfer of spin angular momentum,[13-17] but the oscillation linewidths remain similar up to 20 GHz. In contrast, the damping and coercivity increase by one order of magnitude when growing films at low oxygen pressure due to the presence of vacancies and Gd migration during annealing, forming a ~5 nm thick Fe-Gd mixed GdIG layer at the GGG/YIG interface,[18] see Figs. 1a-b.

The enhanced damping with GdIG is also apparent in Time-Resolved Scanning Kerr Microscopy (TRSKM) measurements. In TRSKM, magnetization dynamics from sub-ps to 10s ns are probed by combining time-resolved spectroscopy, the magneto-optic Kerr effect and microwave excitations of the FMR with a coplanar waveguide.[19,20] When measuring pristine YIG films and YIG/PtMn(5nm) bilayers, a single pulsed microwave excitation generates oscillations that persist in both samples for 10s of ns and remain measurable after 50 precession cycles. However, in GdIG/YIG, where the damping is one order of magnitude higher, the oscillations decay in much shorter timescales (Fig. 1c). GdIG is non-magnetic at room temperature, but ferrimagnetic with a net magnetisation antiparallel to YIG at temperatures below 100-150 K due to an AFM coupling between Fe and Gd. This coupling leads to in-plane exchange bias and increased coercivity at low temperatures (Fig. 1d).

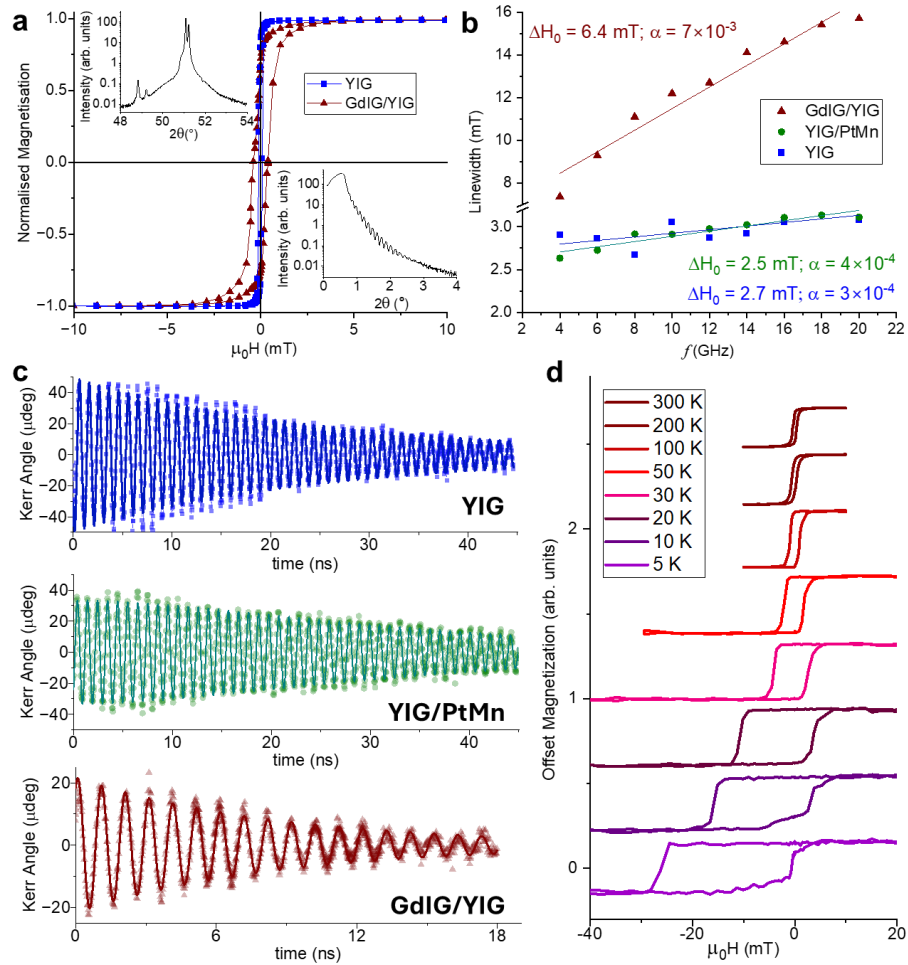

**Figure 1: Characterisation of pristine and interfaced YIG thin films. *a.*** *In-plane hysteresis loops of pristine YIG and GdIG/YIG. The coercivity is increased from <0.1 mT to ~1 mT. Insets: X-ray diffraction and reflectivity of GGG/YIG showing epitaxial growth and low roughness.* ***b.*** *The FMR linewidth and damping are increased by about one order of magnitude in the vacancy-rich, GdIG interfaced YIG film. A PtMn AFM interface deposited on YIG has only a small effect on the linewidth.* ***c.*** *Magnetization precession signals measured from the films via time-resolved scanning Kerr microscopy in a 11.7 mT field show oscillations that can be observed for several 10s of ns in low damping, pristine films, or ~10 ns in GdIG/YIG interfaces. Lines are fits to a stretched cosine function with an exponential decay over characteristic times of 26, 39 and 8.5 ns.* ***d.*** *The AFM coupling at the GdIG interface increases the coercivity and gives rise to an in-plane exchange bias at temperatures ≲ 20 K.*

The coherence of the magnetisation dynamics in YIG films is affected by impurity scattering, causing an increased damping at low temperatures.[21] Magnon diffusion length and the generation of spin currents are also dependent on film thickness and temperature.[15,22] Nevertheless, the FMR frequency and linewidth in pristine YIG remains proportional to an OOP field up to 40 GHz even at 10 K (Fig. 2a). The interfacial exchange coupling generated across FM/AFM interfaces is relatively strong, and a reorientation of the Néel vector can take place in a mechanism that is similar to the 'spin-flop' of conventional antiferromagnets, with a perpendicular orientation between the ferro/antiferromagnetic axis directions.[23] This spin reorientation results in changes to e.g. the spin Seebeck and anomalous Hall effect,[24-26] and can also be observed as a change of sign of the spin Hall magnetoresistance in YIG/PtMn.[27] The spin wave spectrum in AFM-coupled system is different before and after the spin-flop transition, and the magnetisation of the AFM-coupled moments can increase faster than linearly with the magnetic field.[28] The magnetisation of YIG is not pinned by PtMn at 300 K, but at 10 K these changes in the AFM/FM spin alignment affect the YIG magnetisation dynamics. During the realignment of the Néel vector there is an acceleration of the spin dynamics, i.e. a dependence of the resonant frequency as $f \propto H^\beta$ with $\beta = 3$. At high fields ($\mu_0 H \gtrsim 0.5$ T) the FMR returns to a typical $\beta = 1$ dependence from eq. (1). The oscillation damping, defined as the proportionality factor between the linewidth of the microwave absorption and the applied field, is not constant, with an enhanced linewidth during the realignment of the AFM moments at the interface at low fields (Fig. 2b).

GdIG is a compensated ferrimagnet inert at room temperature, but the Gd moment dominates at low temperatures.[18,29] The Gd rare earth ions in GdIG couple antiferromagnetically to the Fe ions in YIG, so the two ferrimagnetic layers, GdIG and YIG, are aligned antiparallel at low fields. This AFM coupling reduces the measured magnetic moment of the bilayer compared to YIG on its own, and it increases the damping further at low temperatures.[30,31] At 10 K, as the magnetic field is increased to some 350 mT, there is a spin-flop, with the GdIG and YIG layers changing from an antiparallel coupling in the field axis to an angled interaction with a different exchange where the moments are not fully antiparallel nor oriented in the direction of the applied field.[32] The result is a discontinuity in the low temperature magnetic resonance, with different $H_{eff}$ at low and high fields. Furthermore, at the spin-flop field there is also a factor 3-5 increase of the oscillation linewidth, see Fig. 2c.

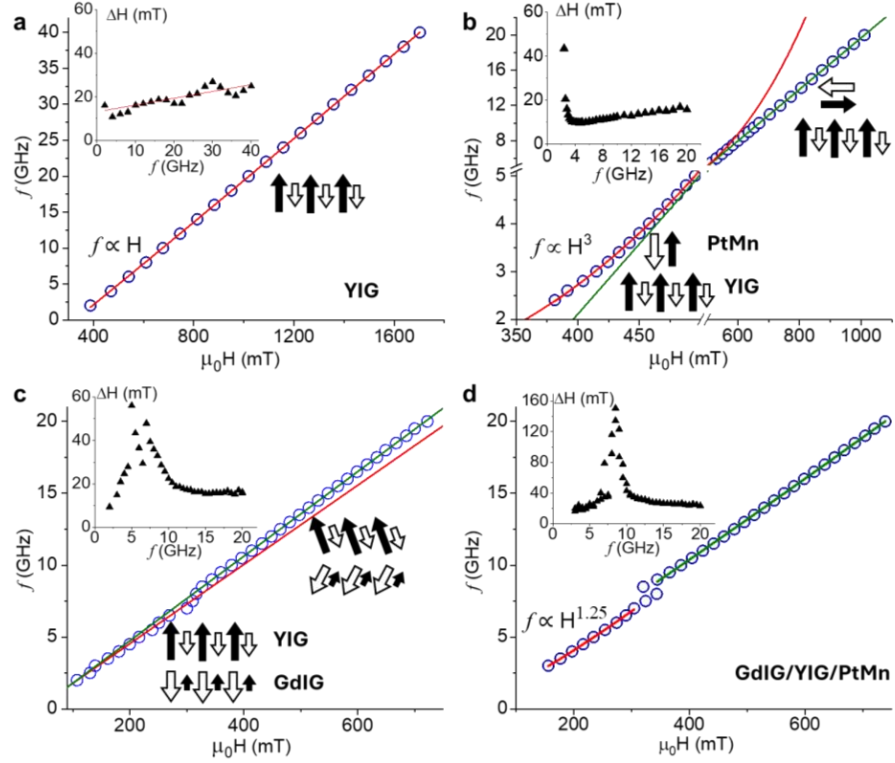

**Figure 2: Low temperature (10 K) OOP FMR measurements for pristine and interfaced YIG films 50-70 nm thick. a.** Pristine YIG shows low damping and a typical linear dependence of frequency vs. field; **b.** In YIG/PtMn, the AFM top interface accelerates the dynamics and increases the linewidth at fields ≲0.5 T. Above 0.5 T, the Neel vector lies perpendicular to the applied field (i.e. in plane) and the PtMn-YIG coupling is suppressed. **c.** FMR for GdIG/YIG. There is a discontinuity in the field dependence at ~300-350 mT due to the spin-flop of the AFM coupled ferrimagnets, leading to a change in anisotropy and a shift of the linear FMR. Around the spin-flop, the resonance linewidth increases by a factor 3-4 due to the reorientation of the magnetic moments w.r.t. the applied field. **d.** Combining both PtMn and GdIG interfaces leads to partly accelerated dynamics at low fields and a heavily suppressed resonance with very strong damping around the spin-flop field. The change in $H_{eff}$ is of the order of 50-100 mT. Lines are linear or power (exponent $\beta$ shown if ≠1) fits of $f$ vs. H at low (red) and high fields (green).

Combining both PtMn and GdIG interfaces results in magnetisation dynamics where the frequency has a power dependence with field; $f \propto H^\beta$ where $1.2 < \beta < 1.5$ at low fields and again $\beta = 1$ at high fields. The linewidth at the spin-flop transition is increased by up to two orders of magnitude and is comparable to the resonance field itself, Fig. 2d. We note that the linewidth for GdIG/YIG/PtMn is not a straightforward superposition of what is observed in each interface separately, but it appears to be dominated by coupling with GdIG leading to a relatively small linewidth at low frequencies and an even larger peak at the spin-flop field.

To explore this linewidth peak observed in GdIG/YIG/PtMn trilayers, we measure the FMR at a range of temperatures and fields. The large damping at low temperatures, when the YIG magnetic moments are coupled to both interfaces, can be attributed to a combination of the reorientation of the Néel vector in PtMn and the spin-flop in GdIG/YIG resulting in a magnetic structure within the YIG film that may be non-collinear and/or not fully oriented in the DC field axis (Fig. 3a). As the temperature is lowered, the FMR signal for a given frequency first increases and moves to higher fields due to the larger magnetisation and power absorption. However, as PtMn blocks the YIG layer, impurity scattering starts to dominate and GdIG couples antiferromagnetically, the magnetisation decreases, the damping increases and the FMR signal is reduced, see Fig. 3b. Below ~20 K, the FMR signal corresponding to the spin realignment fields is quenched, offering a range of microwave frequencies, around 7-10 GHz for our samples, with a very small and broad power absorption, see Figure 3c.

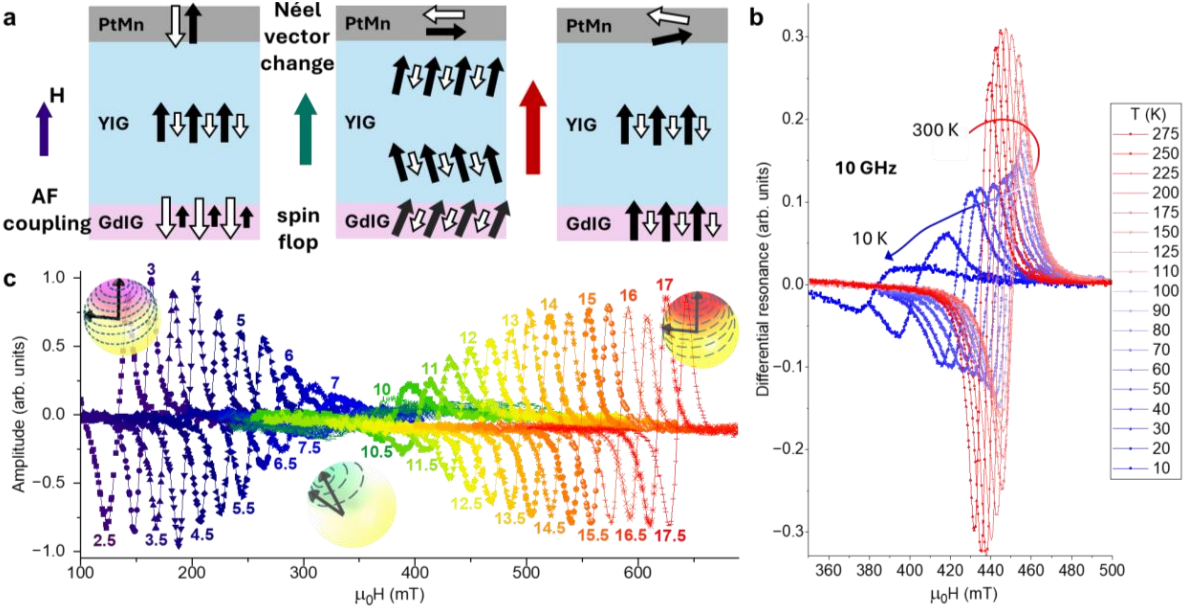

**Figure 3: FMR suppression in double-coupled YIG films.** *a. Schematic of the spin reorientations as a function of the applied field. b. Temperature dependence of the 10 GHz resonance in GdIG/YIG/PtMn. c. Differential FMR measured at 10 K showing the reduced absorption and increased linewidth near the reorientation field. Labels are the frequency of the measurement in GHz. Note that measurements at 7.5-9.5 GHz are included but cannot be distinguished from the background.*

In YIG/PtMn, the $f \propto H^\beta$ dependence with $\beta>1$ and enhanced low-field linewidth emerges as the AFM blocks the YIG layer, and extends until the change of the Néel vector direction is completed, when the frequency and linewidth depend again linearly on the magnetic field, see Figs. 4a&b. Similarly, the quenching of the oscillation in GdIG/YIG/PtMn can be seen to start around 30 K, with an asymptotic increase in the linewidth with decreasing temperature (Fig. 4c). The spin-flop leads then to a peak in the linewidth with frequency around 8-9 GHz

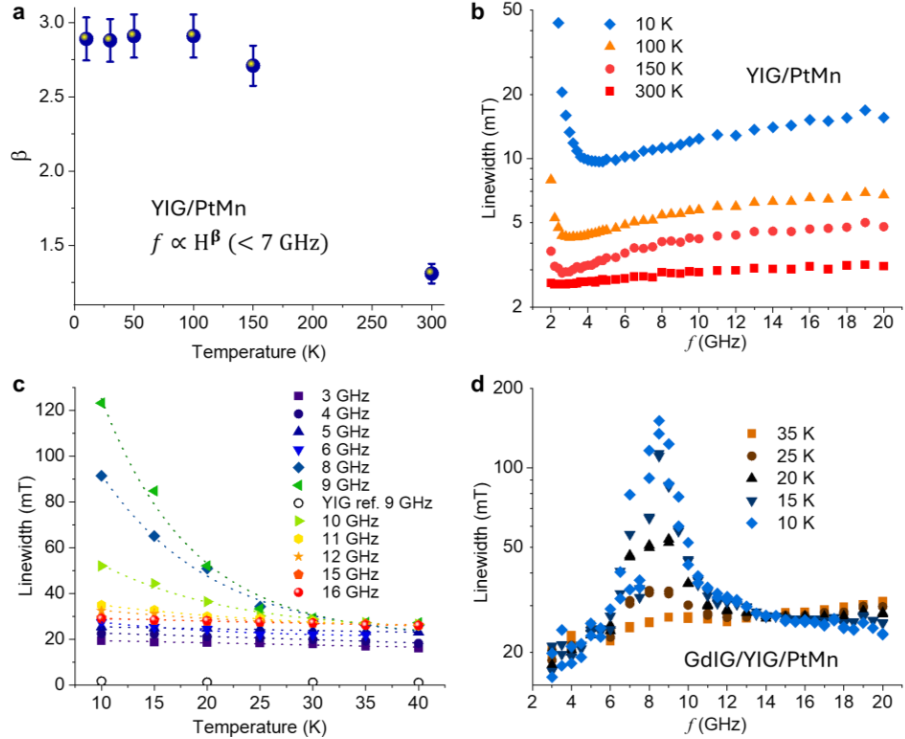

**Figure 4: Temperature dependence of FMR parameters.** *a. Exponent in the low field power dependence for YIG/PtMn samples. b. The linewidth of YIG/PtMn shows the strengthening of the ferri-AFM coupling as the temperature is lowered. c. The peak in the linewidth of GdIG/YIG/PtMn corresponding to the spin flop and FMR suppression around 8-10 GHz emerges below some 30 K, when GdIG is fully magnetised and the PtMn blocks the YIG magnetisation. Lines are fits to an asymptotic function. d. Changes in linewidth as a function of frequency and temperature below 40 K.*

as the temperature is lowered, see Fig. 4d.

The power-law dependence of the frequency with field in AFM/FM interfaces such as YIG/PtMn and shown in Fig. 4a can be controlled by the strength of the exchange coupling, which could be applied in the design of non-linear magnonics converters. Similarly, the suppression of the magnetisation dynamics can be used in the design of notch filters for a range of frequencies at which the magnetic material will not absorb microwave power or generate spin pumping. In GdIG/YIG/PtMn these effects take place below 30 K and in a frequency range of 7-10 GHz corresponding to the resonant frequencies around 350 mT, but this is dependent on the magnetisation of the ferrimagnetic material and the strength of the coupling, so the frequency/field range and temperatures could be tuned by using different materials and/or interfaces.

**METHODS**

**Thin film growth:** YIG films were grown on GGG substrates from $Y_3Fe_5O_{12}$ polycrystalline targets by both off-axis RF sputtering at high temperatures in an oxygen atmosphere and RF reactive sputtering in oxygen atmosphere followed by annealing at 850 $^oC$ in air. In order to obtain a GdIG/YIG interface, YIG films were grown at lower oxygen pressure to generate vacancies in the YIG film for Gd ions to diffuse to.

**Ferromagnetic resonance:** OOP FMR was carried out using NanOsc Instruments inserts and waveguides adapted to a Quantum Design SQUID-VSM MPMS3. The measurements are carried out at a fixed frequency with a varying DC field. Broadband spectroscopy allows for measurements up to 40 GHz modulated by a low frequency AC field of up to several Oe.

**TRSKM:** Time-resolved polar Kerr measurements in response to an in-plane pulsed microwave magnetic field were performed at normal incidence in a scanning microscope. Probing laser pulses were generated from an optical parametric amplifier (OPA) operating at a wavelength of 800 nm and 1 MHz repetition rate. Emerging optical pulses were compressed to ~50 fs before second harmonic generation was used to produce probe pulses with 400 nm wavelength for greater sensitivity to the magnetization in iron garnets. The probe beam was optically delayed by up to 8 ns with sub-ps resolution before being linearly polarised and focused onto the sample using a ×10 microscope objective lens. The 1 MHz laser sync output of the OPA pump laser was used to trigger an electronic delay generator. The trigger was prescaled ×4 to generate an electronically delayed output TTL pulse with repetition rate 250 kHz that was subsequently used to trigger a pulse generator with up to 2 ns pulse duration and 40 V output amplitude. The output was tuned to provide a ~15 V pulse with a 300 ps duration that was passed through a coplanar waveguide to generate the pulsed magnetic field used to excite the sample magnetisation. The waveguide was made on a high frequency printed circuit board onto which the sample was placed face up, while the probe beam was focused over the centre conductor with a spot size smaller than the conductor width. The in-plane equilibrium magnetic state of the sample was set using a permanent magnet assembly mounted on motorised translation and rotation stages to precisely control the field strength and direction. Polar Kerr signals corresponding to the change in the out-of-plane component of the dynamic magnetization were detected using a balanced photodiode polarizing bridge detector with a 1 MHz bandwidth. A high frequency lock-in amplifier was used to recover the dynamic signal at 250 kHz from the 1 MHz probe pulses using a 250 kHz reference signal from the delay generator.

**ACKNOWLEDGEMENTS**

We thank the Engineering and Physical Sciences Research Council in the UK for financial support via the grants EP/S030263/1 and EP/X027074/1. Time-resolved measurements were performed in the Exeter Time-Resolved Magnetism (EXTREMAG) Facility established by grants EP/R008809/1 and EP/V054112/1.

[1] A. Krysztofik, S. Özoglu, R. D. McMichael, and E. Coy, Scientific Reports **11**, 14011 (2021).
[2] C. Kittel, Physical Review **73**, 155 (1948).
[3] O. d'Allivy Kelly *et al.*, Applied Physics Letters **103**, 082408 (2013).
[4] H. C. Chang, P. Li, W. Zhang, T. Liu, A. Hoffmann, L. J. Deng, and M. Z. Wu, Ieee Magnetics Letters **5**, 6700104 (2014).
[5] M. C. Onbasli, A. Kehlberger, D. H. Kim, G. Jakob, M. Kläui, A. V. Chumak, B. Hillebrands, and C. A. Ross, Apl Materials **2**, 106102 (2014).
[6] C. Hauser *et al.*, Scientific Reports **6**, 20827 (2016).
[7] G. Schmidt, C. Hauser, P. Trempler, M. Paleschke, and E. T. Papaioannou, Physica Status Solidi B-Basic Solid State Physics **257**, 1900644 (2020).
[8] Y. C. Yang, T. Liu, L. Bi, and L. J. Deng, Journal of Alloys and Compounds **860**, 158235 (2021).
[9] A. Z. Arsad, A. W. M. Zuhdi, N. B. Ibrahim, and M. A. Hannan, Applied Sciences-Basel **13**, 1218 (2023).

[10] Y. Nakamura, S. B. S. Chauhan, and P. B. Lim, Photonics **11**, 931 (2024).
[11] M. Pardavi-Horvath, Journal of Magnetism and Magnetic Materials **215**, 171 (2000).
[12] V. G. Harris, Ieee Transactions on Magnetics **48**, 1075 (2012).
[13] B. Heinrich, C. Burrowes, E. Montoya, B. Kardasz, E. Girt, Y. Y. Song, Y. Y. Sun, and M. Z. Wu, Physical Review Letters **107**, 066604 (2011).
[14] C. H. Du, H. L. Wang, F. Y. Yang, and P. C. Hammel, Physical Review Applied **1**, 044004 (2014).
[15] Z. Fang, A. Mitra, A. L. Westerman, M. Ali, C. Ciccarelli, O. Cespedes, B. J. Hickey, and A. J. Ferguson, Applied Physics Letters **110**, 092403 (2017).
[16] W. Zhang, M. B. Jungfleisch, W. J. Jiang, J. E. Pearson, A. Hoffmann, F. Freimuth, and Y. Mokrousov, Physical Review Letters **113**, 196602 (2014).
[17] Y. X. Ou, S. J. Shi, D. C. Ralph, and R. A. Buhrman, Physical Review B **93**, 220405 (2016).
[18] A. Mitra *et al.*, Scientific Reports **7**, 11774 (2017).
[19] P. S. Keatley *et al.*, Applied Physics Letters **118**, 122405 (2021).
[20] P. S. Keatley, V. V. Kruglyak, A. Neudert, E. A. Galaktionov, R. J. Hicken, J. R. Childress, and J. A. Katine, Physical Review B **78**, 214412 (2008).
[21] C. L. Jermain, S. V. Aradhya, N. D. Reynolds, R. A. Buhrman, J. T. Brangham, M. R. Page, P. C. Hammel, F. Y. Yang, and D. C. Ralph, Physical Review B **95**, 174411 (2017).
[22] A. Talalaevskij, M. Decker, J. Stigloher, A. Mitra, H. S. Köner, O. Cespedes, C. H. Back, and B. J. Hickey, Physical Review B **95**, 064409 (2017).
[23] N. C. Koon, Physical Review Letters **78**, 4865 (1997).
[24] D. Reitz, J. X. Li, W. Yuan, J. Shi, and Y. Tserkovnyak, Physical Review B **102**, 020408 (2020).
[25] Y. Zhang *et al.*, Nanoscale **12**, 23266 (2020).
[26] R. C. Bhatt, L. X. Ye, N. T. Hai, J. C. Wu, and T. H. Wu, Journal of Magnetism and Magnetic Materials **537**, 168196 (2021).
[27] A. B. Hari Babu Vasili, Satam Alotibi, Mairi McCauley, Timothy Moorsom, Matthew Rogers, Manuel Valvidares, Pierluigi Gargiani, Donald MacLaren, Mannan Ali, Gavin Burnell, Bryan Hickey, David O'Regan, Stefano Sanvito and Oscar Cespedes, Research Square (2025).
[28] Y. L. Wang and H. B. Callen, Journal of Physics and Chemistry of Solids **25**, 1459 (1964).
[29] P. J. Wang, J. T. Ke, G. S. Li, L. Z. Bi, C. Q. Hu, Z. Z. Zhu, J. H. Liu, Y. Zhang, and J. W. Cai, Applied Physics Letters **124**, 172405 (2024).
[30] R. Kumar, B. Samantaray, S. Das, K. Lal, D. Samal, and Z. Hossain, Physical Review B **106**, 054405 (2022).
[31] S. Satapathy, P. K. Siwach, H. K. Singh, R. P. Pant, and K. K. Maurya, Physica B - Condensed Matter **669**, 415278 (2023).
[32] P. Steadman, M. Ali, A. T. Hindmarch, C. H. Marrows, B. J. Hickey, S. Langridge, R. M. Dalgliesh, and S. Foster, Physical Review Letters **89**, 077201 (2002).